\documentclass[english,aps,preprint]{revtex4}
\usepackage[T1]{fontenc}
\usepackage[latin9]{inputenc}
\usepackage{amsmath}
\usepackage{graphicx}
\usepackage{amssymb}

\providecommand{\tabularnewline}{\\}

\usepackage{babel}

\begin{document}

\title{A Search for the Fourth SM Family Quarks through Anomalous Decays}

\author{M. SAHIN}

\email{m.sahin@etu.edu.tr}

\affiliation{TOBB University of Economics and Technology, Physics Division, Ankara,
Turkey}

\author{S. SULTANSOY}

\email{ssultansoy@etu.edu.tr}

\affiliation{TOBB University of Economics and Technology, Physics Division, Ankara,
Turkey}

\affiliation{Institute of Physics, National Academy of Sciences, Baku, Azerbaijan}

\author{S. TURKOZ}

\email{turkoz@science.ankara.edu.tr}

\affiliation{Ankara University, Department of Physics, Ankara, Turkey}
\begin{abstract}
The existence of fourth family follows from the basics of the Standard
Model. Because of the high masses of the fourth family quarks, their
anomalous decays could be dominant, if certain criteria are met. This
will drastically change the search strategy at hadron colliders. We
show that the fourth SM family down quarks with masses up to $400-450$
GeV can be observed (or excluded) via anomalous decays by Tevatron
before the LHC. 
\end{abstract}
\maketitle
Recently, the search for the fourth SM family has drawn attention
of both theoretical and experimental HEP community \cite{Holdom,Beyond 3 SM CERN,Beyond 3 SM Taiwan}.
The existence of the fourth SM family is the outcome of the flavor
democracy hypothesis and the actual spectrum of the third family fermion
masses \cite{Fritzsch1,Datta,Celikel} (see, also review \cite{Sultansoy}
and ref's therein). Twenty years' ongoing discussions on {}``precision
data vs fourth SM family'' have been concluded in favor of the fourth
SM family (see \cite{OPUCEM,Eberhardt} and ref's therein). Today,
the mass and the mixing patterns of quarks and leptons are the most
mysterious aspects of particle physics. Within the SM all these masses
and mixings are put by hand and constitute the basic source of parameter
inflation. The discovery of the fourth SM family could provide some
systematics for SM fermion masses and mixings, which seems chaotic
in the three family case.

Let us remind that flavour physics met a lot of surprises. The first
example was discovery of $\mu$-meson (We were looking for $\pi$-meson
predicted by Yukawa but discovered the {}``heavy electron''). The
next example was represented by strange particles (later we understood
that they contains strange quarks). The story was followed by $\tau$-lepton,
c- and b-quarks discovered in 1970's. Actually, c-quark was foreseen
by GIM mechanism and quark-lepton symmetry and its mass was estimated
in the few GeV region, whereas the discovery of $\tau$-lepton and
b-quark was completely surprising for physicists. According to the
Standard Model they are the members of the third fermion family, which
was completed by the discovery of t-quark in 1995 at Tevatron. Actually,
we need at least three fermion families in order to handle CP-violation
within the SM \cite{Kobayashi}. CP violation is necessary for the
explanation of Barion Asymmetry of the Universe (BAU). Unfortunately,
SM with three fermion families does not provide actual magnitude for
BAU. Fortunately, the fourth SM family could provide additional factor
of order of $10^{10}$ and, therefore, solve the problem \cite{Hou}.

The existence of the fourth SM family could affect the Tevatron physics
search program in two ways. First, it leads to essential enhancement
of the Higgs production via gluon fusion (see, \cite{E. Arik,N.Becerici}
and ref's therein). As a result, while in the case of three SM families
(SM3) Tevatron data on $gg\rightarrow H\rightarrow WW\rightarrow ll^{'}$
$P_{T}^{mis}$ process excludes $162$ GeV$<M_{H}$$<$$166$ GeV
\cite{higgs SM3} at $95\%$ CL, the excluded region becomes $131$
GeV $<M_{H}<204$ GeV \cite{Higgs SM4}, if the Nature prefers four
SM families case. Second, if the masses of the fourth family quarks
are not so large, they could be directly observed by Tevatron.

Up to now, all searches for the fourth family quarks have been done
by assuming that SM decay modes are dominant. Current lower limits
on the fourth SM family quark masses from direct searches at the Tevatron
are:

a) $M_{d_{4}}>338$ GeV at $95\%$ CL coming from the search for new
bottom-like quark pair decays in same-charged lepton events with an
integrated luminosity of $2.7$ fb$^{-1}$ \cite{Aaltonen1},

b) $M_{u_{4}}>256$ GeV at $95\%$ CL coming from the search for heavy
top-like quarks, using lepton plus jets events with an integrated
luminosity $0.76$ fb$^{-1}$\cite{Aaltonen2}.

The heaviness of the t-quark has induced searches for its anomalous
interactions \cite{Fritzsch2,Abe,Abazov,Aaltonen3,Aaltonen4}. For
the same reason \textendash{} new quarks are heavier than t-quark
\textendash{} in the search strategy the anomalous interactions
for the fourth SM family quarks should be taken into account. These
interactions may manifest themselves both in the production and decays
of new quarks. If the anomalous decay modes of the fourth family quarks
are dominant, lower limits on their masses given above are not valid
and the search strategy should be changed drastically.

In this letter, we consider pair production of the fourth SM family
down-type quarks with subsequent anomalous decays into photon+jet
and jet+jet channels. More general consideration, including other
manifestations, will be presented in the following publication \cite{Sahin}.

The effective Lagrangian for anomalous magnetic type interactions
of the fourth family quarks is given as \cite{Arik,Cabibbo,Hagiwara}:

\begin{equation}
L={\displaystyle \underset{{\scriptstyle q_{i}}}{{\displaystyle \sum}}\frac{\kappa_{\gamma}^{q_{i}}}{\Lambda}e_{q}g_{e}\bar{q_{4}}\sigma_{\mu\nu}q_{i}F^{\mu\nu}+\underset{{\scriptstyle q_{i}}}{{\displaystyle \sum}}\frac{\kappa_{Z}^{q_{i}}}{2\Lambda}}g_{Z}\bar{q_{4}}\sigma_{\mu\nu}q_{i}Z^{\mu\nu}+\underset{{\scriptstyle q_{i}}}{{\displaystyle \sum}}\frac{{\displaystyle \kappa_{g}^{q_{i}}}}{\Lambda}g_{s}\bar{q_{4}}\sigma_{\mu\nu}T^{a}q_{i}G_{a}^{\mu\nu}+H.c.\end{equation}

\hfill{}\hfill{}

where $F^{\mu\nu}$, $Z^{\mu\nu}$ and $G^{\mu\nu}$ are the field
strength tensors of the gauge bosons, $\sigma_{\mu\nu}$ is the anti-symmetric
tensor, $T^{a}$ are Gell-Mann matrices, $e_{q}$ is electric charge
of quark, $g_{e},$ $g_{Z}$ and $g_{s}$ are electromagnetic, neutral
weak and strong coupling constants, respectively. $g_{Z}=g_{e}/cos\theta_{W}sin\theta_{W}$
where $\theta_{W}$ is the Weinberg angle. $\kappa_{\gamma}$, $\kappa_{Z}$
and $\kappa_{g}$ are the strength of anomalous couplings with photon,
$Z$ boson and gluon, respectively. $\Lambda$ is the cutoff scale
for new physics. This type of gauge and Lorentz invariant effective
Lagrangian have been proposed in the framework of composite models
for interactions of excited fermions with ordinary fermions and gauge
bosons \cite{Cabibbo,Hagiwara} . 

For numerical calculations we implement the Lagrangian (1), as well
as fourth family SM Lagrangian into the CalcHEP package \cite{Pukhov}.
The partial decay widths of $d_{4}$ for SM ($d_{4}\rightarrow W^{-}q$
where $q=u$, $c$, $t$) and anomalous ($d_{4}\rightarrow\gamma q$,
$d_{4}\rightarrow Zq$, $d_{4}\rightarrow gq$ where $q=d$, $s$,
$b$) modes are given below:

\begin{equation}
\Gamma(d_{4}\rightarrow W^{-}q)=\frac{|V_{qd_{4}}|^{2}\alpha_{e}M_{d_{4}}^{3}}{16M_{{\scriptscriptstyle W}}^{2}sin^{2}\theta_{{\scriptscriptstyle W}}}\chi_{{\scriptscriptstyle W}}\sqrt{\chi_{{\scriptscriptstyle 0}}}\end{equation}

\medskip{}

where $\chi_{{\scriptstyle {\scriptscriptstyle W}}}=(1+x_{q}^{4}+x_{q}^{2}x_{{\scriptscriptstyle W}}^{2}-2x_{q}^{2}-2x_{{\scriptscriptstyle W}}^{4}+x_{{\scriptscriptstyle W}}^{2})$,
$\chi_{0}=(1+x_{W}^{4}+x_{q}^{4}-2x_{W}^{2}-2x_{q}^{2}-2x_{W}^{2}x_{q}^{2})$,
$x_{q}=(M_{q}/M_{d_{4}})$ and $x_{{\scriptscriptstyle W}}=(M_{{\scriptscriptstyle W}}/M_{d_{4}})$,

\begin{equation}
\Gamma(d_{4}\rightarrow Zq)=\frac{\alpha_{e}M_{d_{4}}^{3}}{16cos^{2}\theta_{{\scriptscriptstyle W}}sin^{2}\theta_{{\scriptscriptstyle W}}}(\frac{\kappa_{{\scriptscriptstyle Z}}^{q}}{\Lambda})^{2}\chi_{{\scriptscriptstyle Z}}\sqrt{\chi_{{\scriptscriptstyle 1}}}\end{equation}

\medskip{}

where $\chi_{Z}=(2-x_{Z}^{4}-x_{Z}^{2}-4x_{q}^{2}-x_{q}^{2}x_{Z}^{2}-6x_{q}x_{Z}^{2}+2x_{q}^{4})$,
$\chi_{1}=(1+x_{Z}^{4}+x_{q}^{2}-2x_{Z}^{2}-2x_{q}^{2}-2x_{Z}^{2}x_{q}^{2})$
and $x_{Z}=(M_{{\scriptscriptstyle Z}}/M_{d_{4}})$

\begin{equation}
\Gamma(d_{4}\rightarrow gq)=\frac{2\alpha_{s}M_{d_{4}}^{3}}{3}(\frac{\kappa_{g}^{q}}{\Lambda})^{2}\chi_{{\scriptscriptstyle 2}}\end{equation}

\medskip{}

where $\chi_{2}=(1-3x_{q}^{2}+3x_{q}^{4}-x_{q}^{6})$,

\begin{equation}
\Gamma(d_{4}\rightarrow\gamma q)=\frac{\alpha_{e}M_{d_{4}}^{3}Q_{q}^{2}}{2}(\frac{\kappa_{\gamma}^{q}}{\Lambda})^{2}\chi_{{\scriptscriptstyle 2}}\end{equation}

\medskip{}

One can wonder what is the criteria for the dominance of anomalous
decay modes over SM ones. It is seen from Eq. (2)-(5) that the anomalous
decay modes of the fourth SM family quarks are dominant, i.e. $\Gamma(d_{4}\rightarrow gq)+\Gamma(d_{4}\rightarrow Zq)+\Gamma(d_{4}\rightarrow\gamma q)>\Gamma(d_{4}\rightarrow W^{-}q)$,
if the relation $(\kappa/\Lambda)\gtrsim1.2(V_{ud_{4}}^{2}+V_{cd_{4}}^{2}+V_{td_{4}}^{2})^{1/2}$
TeV$^{-1}$ is satisfied (hereafter $\kappa_{Z}^{q}=\kappa_{g}^{q}=\kappa_{\gamma}^{q}=\kappa$
is assumed). The experimental upper bounds for the fourth family quark
CKM matrix elements are $|V_{u_{4}d}|\leq0.063$, $|V_{u_{4}s}|\leq0.46$,
$|V_{u_{4}b}|\leq0.47$, $|V_{ud_{4}}|\leq0.044$, $|V_{cd_{4}}|\leq0.46$,
$|V_{td_{4}}|\leq0.47$ \cite{Ozcan}. On the other hand, the predicted
values of these matrix elements are expected to be rather small in
the framework of flavor democracy hypothesis. For example, the mass
matrix parametrization proposed in \cite{Ciftci}, which gives correct
predictions for CKM and MNS mixing matrix elements through use of
SM fermion mass values as input, predicts $|V_{u_{4}d}|=0.0005$,
$|V_{u_{4}s}|=0.0011$, $|V_{u_{4}b}|=0.0014$, $|V_{ud_{4}}|=0.0002$,
$|V_{cd_{4}}|=0.0012$, $|V_{td_{4}}|=0.0014$. In this case, the
anomalous decay modes are dominant, if $(\kappa/\Lambda)>0.0022$
TeV$^{-1}$. The latter correspondens to upper limit $500$ TeV for
new physics scale $\Lambda$, assuming $\kappa=O(1)$. 

The cross-section for the $d_{4}\bar{d_{4}}$ pair production at the
Tevatron is shown in Fig. 1. We have used CalcHEP \cite{Pukhov} with
CTEQ6L \cite{CTEQ6L_Pumplin_Stump} parton distribution functions
for numerical calculations. One can see that if $M_{d_{4}}$ is about
$400$ GeV the Tevatron with $L_{int}=10$ fb$^{-1}$ will yield more
than hundred $d_{4}\bar{d_{4}}$ pairs.

\begin{figure}
\includegraphics[scale=0.7]{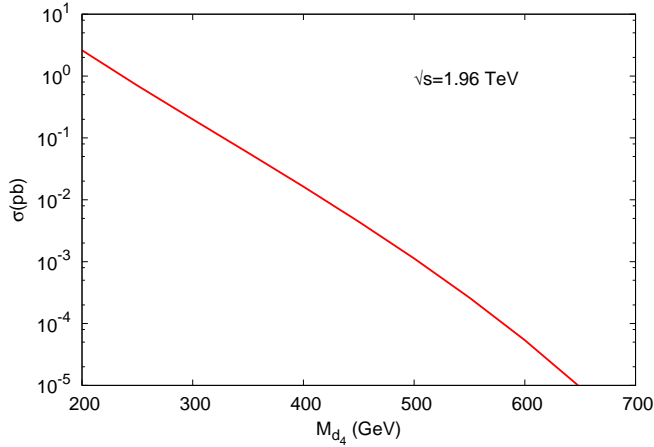}

\caption{Cross section for $d_{4}\bar{d_{4}}$ pair production at the Tevatron.}

\end{figure}

We propose $p\bar{p}$$\rightarrow d_{4}\bar{d_{4}}$$\rightarrow\gamma qg\overline{q}$
(where $q=d,s,b$) process to analyze the Tevatron search potential
to discover $d_{4}$ quark via anomalous decays. In detector this
process is seen as $\gamma+3j$ events. We use CalcHEP \cite{Pukhov}
and MADGRAPH \cite{Maltoni} packages with CTEQ6L \cite{CTEQ6L_Pumplin_Stump}
parton distribution functions for the calculation of the signal and
background processes, respectively. In order to extract the $d_{4}$
signal and to suppress the background, following cuts are applied:
$p_{T}>50$ GeV and $|\eta|<2$ for all final state partons and photon,
as well as invariant mass within $\pm$ $20$ GeV around $d_{4}$
mass. The effects of these cuts can be seen from Table I.

\begin{table}
\begin{tabular}{|c|c|c|c|c|c|c|}
\hline 
$M_{d_{4}}$  & \multicolumn{2}{c|}{$200$ GeV} & \multicolumn{2}{c|}{$300$ GeV} & \multicolumn{2}{c|}{$400$ GeV}\tabularnewline
\hline
\hline 
cuts  & $\sigma_{S}$, fb  & $\sigma_{B}$, fb  & $\sigma_{S}$, fb  & $\sigma_{B}$, fb  & $\sigma_{S}$, fb  & $\sigma_{B}$, fb\tabularnewline
\hline 
$p_{T}>20$GeV  & $39.2$  & $5.4\times10^{5}$  & $2.92$  & $5.4\times10^{5}$  & $0.23$  & $5.4\times10^{5}$\tabularnewline
\hline 
$p_{T}>50$GeV  & $24.5$  & $2.7\times10^{3}$  & $2.40$  & $2.7\times10^{3}$  & $0.21$  & $2.7\times10^{3}$\tabularnewline
\hline 
all cuts  & $21.8$  & $3.63$  & $2.27$  & $0.091$  & $0.20$  & $0.006$\tabularnewline
\hline
\end{tabular}

\caption{Signal and background cross sections values for various cuts. All
cuts include $p_{T}>50$ GeV, $|\eta|<2$, $|M_{inv}(\gamma j)-M_{d_{4}}|<20$
GeV, $|M_{inv}(jj)-M_{d_{4}}|<20$ GeV.}

\end{table}

Statistical significance has been calculated by using following formula
\cite{CMS_Note_significance}:

\begin{equation}
S=\sqrt{2[(s+b)ln(1+\frac{s}{b})-s]}\end{equation}

where $s$ and $b$ represents the numbers of signal and background
events, respectively.

In Fig. 2 the necessary integrated luminosities for the observation
of $d_{4}$ quark at the Tevatron are plotted as a function of $d_{4}$
mass. It is seen that the fourth family down quarks with masses below
375 GeV could be excluded at $95\%$ CL with $2.7$ fb$^{-1}$ integrated
luminosity, if the corresponding analysis of data has been performed
(compare with 338 GeV for SM modes dominant case \cite{Aaltonen1}).
Discovery, observation and exclusion for the different values of the
Tevatron integrated luminosity are given in Table II.

\begin{figure}
\includegraphics[scale=0.7]{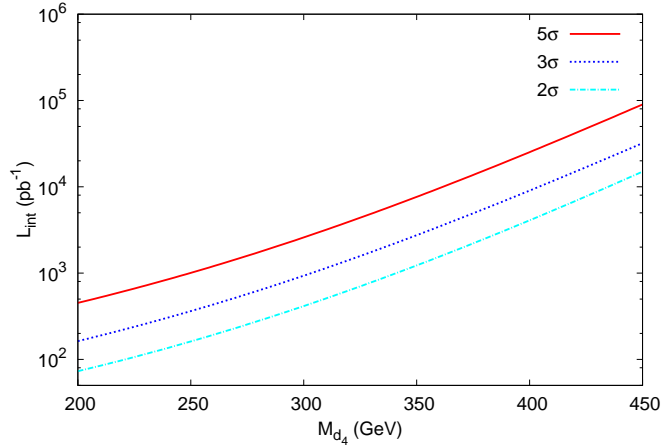}

\caption{The necessary integrated luminosity for the observation of $d_{4}$
quark at the Tevatron}

\end{figure}

\begin{table}
\begin{tabular}{|c|c|c|c|}
\hline 
$L_{int}$, fb$^{-1}$  & $5$  & $10$  & $20$\tabularnewline
\hline
\hline 
$2\sigma$ exclusion  & $390$ GeV  & $430$ GeV  & $460$ GeV\tabularnewline
\hline 
$3\sigma$ observation  & $370$ GeV  & $410$ GeV  & $440$ GeV\tabularnewline
\hline 
$5\sigma$ discovery  & $340$ GeV  & $360$ GeV  & $390$ GeV\tabularnewline
\hline
\end{tabular}

\caption{Reachable $M_{d_{4}}$ mass values for discovery, observation and
exclusion at the Tevatron. }

\end{table}

\emph{Summary and Outlook}. \textendash{} Keeping in mind the LHC
status, the Tevatron has about 2 more years for new physics discovery.
The fourth SM family quarks are among the most prominent candidates
for beyond the SM3 physics and possible dominance of their anomalous
decay modes should not be ignored. If these modes are dominant, the
fourth SM family down quarks with masses up to 400-450 GeV can be
observed (or excluded) via anomalous decays by Tevatron before the
LHC. 
\begin{acknowledgments}
This work is supported by DPT, TAEK and TUBITAK. \end{acknowledgments}

\end{document}